\documentclass{aa}  

\usepackage{graphicx}
\usepackage{txfonts}
\newcommand{\nc}{\newcommand}
\def\cplus{C$^{+}$}
\nc{\Msun}{\ensuremath{\mathrm{M}_\odot}}
\nc{\Rsun}{\ensuremath{\mathrm{R}_\odot}}
\nc{\cmcub}{\mbox{cm$^{-3}$}}
\newcommand\arcdeg{\mbox{$^\circ$}}%  
\newcommand{\CII}{[C {\sc ii}]}
\newcommand{\thCII}{[$^{13}$C {\sc ii}]}
\newcommand{\NII}{[N {\sc ii}]}

\nc\micron{\mbox{$\mu$m}}
\newcommand{\OI}{[O {\sc i}]}
\nc{\cmsq}{\mbox{cm$^{-2}$}}
\nc{\kms}{\mbox{km~s$^{-1}$}}
\nc{\lsun}{\ensuremath{\mathrm{L}_\odot}}
\newcommand{\HII}{H {\sc ii}}

\begin{document}

   \title{\cplus\ distribution around S\,1 in
   $\rho$\,Ophiuchi}

   \author{B. Mookerjea \inst{1} 
         \and G. Sandell \inst{2,3}
          \and
          W. Vacca \inst{3}
          \and
          E. Chambers \inst{3}
          \and  R. G\"usten \inst{4}
          }

   \institute{Tata Institute of Fundamental Research, Homi Bhabha Road,
Mumbai 400005, India
              \email{bhaswati@tifr.res.in}
         \and
         Institute for Astronomy, University of Hawaii,  640 N. Aohoku Place, Hilo, HI 96720, USA
\and USRA/SOFIA, NASA Ames Research Center, Mail Stop 232-12, Building N232, P.O. Box 1, Moffett Field, CA 94035-0001, USA
         \and
Max Planck Institut f\"ur Radioastronomie, Auf dem H\"ugel 69, 53121 Bonn, Germany     
             }

   \date{latest revision \today}

% \abstract{}{}{}{}{} 
% 5 {} token are mandatory
\abstract
{We analyze a \CII\ 158\,\micron\ map obtained with the L2 GREAT receiver on
SOFIA of the emission/reflection nebula illuminated by the early B star S\,1  in
the $\rho$\,Oph\,A cloud core. This data set has been complemented with maps of 
CO(3--2), $^{13}$CO(3--2) and C$^{18}$O(3-2), observed as a part of the JCMT
Gould Belt Survey, with archival HCO$^+$(4--3) JCMT data, as well as with \OI\
63 and 145\,\micron\ imaging with {\it Herschel}/PACS.  The \CII\ emission is
completely dominated by the strong PDR emission from the nebula surrounding S\,1
expanding into the dense Oph\,A molecular cloud west and south of S\,1. The
\CII\ emission is significantly blue shifted relative to the CO spectra
and also relative to the systemic velocity, particularly in the northwestern 
part of the nebula.
The \CII\ lines are broader towards the center of the S\,1 nebula and narrower
towards the PDR shell. 
The \CII\ lines are strongly self-absorbed over an extended region in
the S\,1 PDR. Based on the strength of the \thCII\ F = 2--1
hyperfine component, \CII\ is significantly optically thick over most of
the nebula. CO and $^{13}$CO(3--2) spectra are strongly self-absorbed,
while C$^{18}$O(3--2) is single peaked and centered in the middle of the
self-absorption. We have used a simple two-layer LTE model to
characterize the background and foreground cloud contributing to the
\CII\ emission. From this analysis we estimate the extinction due to the
foreground cloud to be $\sim$ 9.9\,mag, which is slightly less than the
reddening estimated towards S\,1.  Since some of the hot gas in the PDR
is not traced by low J CO emission, this result appears quite plausible.
Using a plane parallel PDR model with the observed \OI(145)/\CII\
brightness ratio and an estimated FUV intensity of 3100--5000\,G$_0$
suggests that the density of the \CII\ emitting gas is $\sim$ 3 --
4$\times10^3$\,\cmcub.}

%  \keywords{try }
\keywords{ISM: Clouds -- Submillimeter:~ISM -- ISM: lines and bands
-- ISM: individual ($\rho$\,Oph\,A)  -- ISM: molecules  -- \HII\ regions }

\maketitle
%
%-------------------------------------------------------------------

\section{Introduction}

Photon Dominated Regions (PDRs)  are regions where FUV
(6\,eV $<$ h$\nu <$ 13.6 eV) radiation from young massive stars dominate the
physics and the chemistry of the interstellar medium \citep{Tielens85}.
The PDRs play an important role in reprocessing much of the energy from
stars and re-emitting this energy in the infrared-millimeter
wavelengths. Most of the mass of the gas and dust in the Galaxy resides
in PDRs \citep{Hollenbach99}. In the far infrared the most important
cooling lines are the fine structure lines of \CII\ at 158\,\micron, and
\OI\ at 63 \& 145\,\micron\ and to a lesser extent high-$J$ CO lines,
while PAH emission and H$_2$ lines dominate in the near- and mid-IR.

The $\rho$ Ophiuchi dark cloud complex is the nearest low- and
intermediate-mass star forming region (see \citet{Wilking08} for a
recent review).  The mean distance to  $\rho$ Oph is 137.3 $\pm$ 1.2 pc
based on trigonometric parallax observations with the  Very Long
Baseline Array (VLBA), while the eastern streamer appears to be  $\sim$
10 pc  further away \citep{Ortiz17}. The most massive cores in $\rho$
Oph dark cloud complex are $\rho$\,Oph\,A \citep{Loren90} in L1688 and
L1689\,N,  each of which is connected with a filamentary system of
streamers extending to the north-east over tens of parsecs
\citep[e.g.,][]{Loren89}.  While only little star formation activity is
observed in the streamers, the westernmost core, $\rho$ Oph\,A, harbors
a rich cluster of young stellar objects (YSOs) ranging from young Class
0 protostars to more evolved Class III objects and is distinguished by a
high star-formation efficiency.  \citet{Yui93} did an extended \CII\ map
with 15\arcmin\ spatial resolution of the  $\rho$ Ophiuchi dark cloud
using the balloon-borne telescope BICE. They found extended \CII\
emission throughout the cloud (8 pc $\times$ 6 pc) with the peak
emission centered on the highly reddened B2 V star HD\,147889 with
perhaps minor contributions from the embedded B stars S\,1 and SR\,3.

Recently \citet{larsson2017} have studied the $\rho$\,Oph\,A region with
HIFI and PACS on the {\it Herschel Space Observatory} together with
archival {\it ISO} ISOCAM-CVF data.  Their spectrophotometric ISOCAM
observations of pure rotational lines of H$_2$ and polycyclic aromatic
hydrocarbon (PAH) molecules  outline a clear spherical shell structure
of warm PDR gas around the embedded  star S\,1 with a radius of $\sim$
80\arcsec.  The observations of the \OI\ lines at 63 and 145\,\micron\
using PACS show that not only do the \OI\ lines mark the boundary seen
in H$_2$, but also show significant emission throughout the region
around the exciting star S\,1.  The map of the line flux ratio
$F_{63\,\mu\,{\rm m}}/F_{145\,\mu\,{\rm m}}$ shows remarkably low values
in some regions, which suggests the presence of a cold foreground cloud
that absorbs most of the $^3$P$_{1}$--$^3$P$_{2}$ 63\,\micron\ radiation
but leaves the higher level $^3$P$_{0}$--$^3$P$_{1}$ 145\,\micron\ line
unaffected. All the existing observations of the PDR tracers of the S\,1
PDR unfortunately lack velocity information, hence the interplay between
the cold foreground cloud and the warmer background PDR material is
still unexplored.

In this paper we present velocity resolved \CII\ maps of an area around
the S\,1 PDR in $\rho$\,Oph\,A. Here we aim to study the emission from
the S\,1 PDR to (a) understand the impact of the PDR if any on the
molecular cloud to the west and (b) detect and characterize the
foreground absorbing cloud observed in \OI.

\section{Datasets}

\subsection{SOFIA}

We have retrieved observations  of the
$^2P_{3/2}$$\rightarrow$$^2P_{1/2}$ fine structure transition of ionized
carbon (\cplus) at 1900.5369\,GHz (157.74\,\micron)  and the
$^3P_1$$\rightarrow$$^3P_{0}$ \NII\ line at 1461.1338 GHz (205 $\mu$m),
of  $\rho$\,Oph\,A from the data archive of the Stratospheric
Observatory for Infrared Astronomy \citep[SOFIA][]{young2012}.  The
observations (PI: Di Li) were done using the German REceiver for
Astronomy at Terahertz frequencies \citep[GREAT;][]{heyminck2012} on
July 16 and 20, 2011.  The beam size for \CII\ was 15\farcs3 and the
velocity resolution before smoothing was 0.029\,\kms,  while the
beamsize for \NII\ was 21\arcsec.  The \CII\ map extends over a
220\arcsec$\times$200\arcsec\ area and was done in total power
on-the-fly mode. The map was divided into five sub-maps, each scanned in
Dec, and sampled every 8\arcsec, with a step size between scans of
8\arcsec.  The map was centered on ($\alpha$, $\delta$) = (16$^{\rm h}$
26$^{\rm m}$ 27\fs60, -24\degr23\arcmin56\farcs3) (J2000) using an
off-position at 16$^{\rm h}$ 29$^{\rm m}$ 49s0,
-24\degr29\arcmin31\farcs0 (J2000). We used a main beam efficiency of
0.51  and 0.54 \citep{heyminck2012} to convert the observed antenna
temperatures to main beam brightness temperature ($T_{\rm mb}$) for
\CII\ and \NII,  respectively. \NII\ was not detected anywhere in the
map. 

Finally we retrieved fully processed  level 3 {\it SOFIA}/FORCAST images at
11.1, 19.7, 31.5, and 37.1 $\mu$m \citep{Herter12}. These were taken in dichroic
mode on 2014-03-29 covering the S\,1 field (Program ID: 02\_0070/Lee Mundy). 
For a description of the FORCAST data reduction pipeline, see \citet{Herter13}.

\subsection{JCMT Gould Belt Survey and ISOCAM data}

For comparison with the \CII\ data, we have used maps of the $J$=3--2
transition of CO, $^{13}$CO and C$^{18}$O \citep{white2015} and $J$=4--3
transition of HCO$^+$. The CO (and its isotopes) datacubes were kindly
provided to us by Dr.  Emily Daubrek-Maunder.  The observations were
taken using the Heterodyne Array Receiver Program (HARP)\citep{Buckle09}
on the James Clerk Maxwell Telescope (JCMT)\footnote{The James Clerk
Maxwell Telescope is operated by the East Asian Observatory on behalf of
The National Astronomical Observatory of Japan; Academia Sinica
Institute of Astronomy and Astrophysics; the Korea Astronomy and Space
Science Institute; the Operation, Maintenance and Upgrading Fund for
Astronomical Telescopes and Facility Instruments, budgeted from the
Ministry of Finance (MOF) of China and administrated by the Chinese
Academy of Sciences (CAS), as well as the National Key R\&D Program of
China (No. 2017YFA0402700).  Additional funding support is provided by
the Science and Technology Facilities Council of the United Kingdom and
participating universities in the United Kingdom and Canada.}, Hawaii.
The C$^{18}$O, $^{13}$CO and CO $J$ = 3--2 rest frequencies are
329.3305525, 330.5879601 and 345.7959899\,GHz, respectively.  HARP has a
beamsize of 14\arcsec\ at 345\,GHz.  The CO(3--2) data have a spectral
resolution of 1 \kms, while the C$^{18}$O and $^{13}$CO data have a
resolution of 0.1 \kms. The HCO$^+$(4--3) data set, corresponding to the
proposal M11AU13, was downloaded directly from the JCMT archive at the
Canadian Astronomical Data Centre (CADC).

We have also retrieved fully processed ISOCAM image cubes (CAM1 \& CAM4)
including pipeline processed photometry of S\,1 from the {\it ISO} Data Archive \citep{Lorente06}.

\section{FUV sources in the $\rho$\,Oph\,A  cloud}

\begin{figure}
\centering

\includegraphics[width=0.48\textwidth]{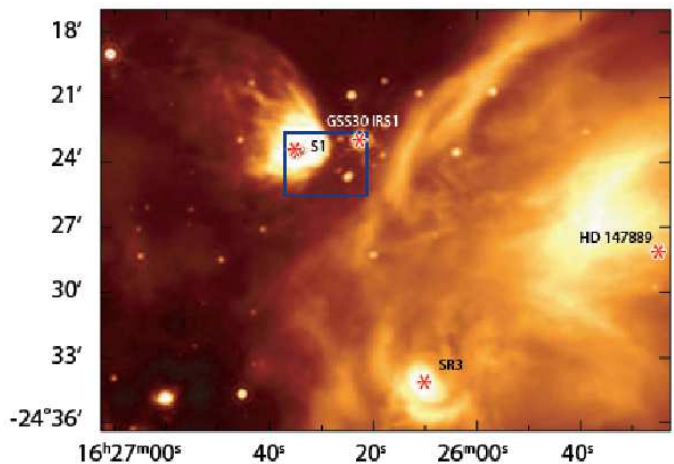}
\caption{Part of MIPS 24 $\mu$m image extracted from Fig. 3 in \citet{Padgett08}. 
The image shows the four main FUV sources in the northern part of 
$\rho$\,Oph: S\,1, GSS\,30\,IRS1, SR\,3 and HD\,147889, which are labeled and marked with
red symbols. The blue rectangle shows the area mapped in \CII\ with GREAT.
\label{fig-MIPS}}
\end{figure}

The $\rho$\,Oph\,A  cloud core is  between the large reflection nebula
illuminated by the B2 V star HD\,147889 on the western side and the
\HII\ region illuminated by the early B star S\,1 on the eastern side
with the embedded B9 - A0 V star SR\,3  \citep{lada1984} to the south,
(see Fig.\,\ref{fig-MIPS}). In the $\rho$\,Oph\,A  core region there is
another bright  star $\sim$175\arcsec\ west of S\,1. This bright mid-IR
source is GSS\,30~IRS\,1 (2MASS J16262138-2423040), a low mass Class I
source illuminating a small bipolar reflection nebula
\citep{Weintraub93}. In the mm/sub-mm  $\rho$\,Oph\,A stands out as a
north-south ridge of several bright dust condensations curving to the
south-east at the southern end \citep{Andre93,motte1998,Liseau15}. One
of these sub-mm sources is VLA\,1623, the prototypical Class 0 source
\citep{Andre93}, which drives a large, highly collimated jet in the
NW-SE direction.

%S1, GY70
{\bf S\,1:} S\,1 is an embedded early B star in the $\rho$\,Oph\,A  cloud at a
distance of 138$\pm$1.7 pc \citep{Ortiz17}. It is the brightest near-IR source
in $\rho$ Ophiuchus \citep{Grasdalen73} and a strong X-ray source
\citep{Montmerle83}. It  was first detected as a radio source in $\rho$ Oph by
\citet{Brown75} (BZ \#4), who suggested that it was a compact \HII\ region
ionized by a B3 star. A detailed VLA study by \citet{Andre88} at 5 and 15 GHz in
the C/D hybrid configuration showed that the radio emission originates from a
nonthermal unresolved core surrounded by a thermal extended halo. They inferred
that the compact nonthermal emission is gyrosynchrotron emission from an
extended magnetosphere around the star, while the the halo is free-free emission
from an optically thin \HII\ region.  From the observed free-free emission, 4
mJy, at 5 GHz, and a half power radius of $\sim$ 10\arcsec, they derived a
spectral type of B3.5 (ZAMS) with an effective temperature of T$_{eff}$ = 17,000
K. The multi-epoch VLBI trigonometric parallax study \citep{Ortiz17} resolves
S\,1 into a binary with a separation of $\sim$ 20 -- 30 mas consistent with
near-IR lunar occultations by \citet{Richichi94,Simon95}, who found S\,1 to have
a late type secondary with a projected distance of 20 mas. From an orbit
analysis of the VLBA data \citet{Ortiz17} derive a mass of 5.8~\Msun\ for the
primary and 1.2 \Msun{} for the secondary. These mass estimates correspond
roughly to B4V and F5V spectral types.
 
S\,1 is too faint in the optical (V = 16.5 mag) to allow accurate
spectral classification. \citet{Cohen79} did not detect H$\alpha$ at all
and assigned it a spectral class of B2. Later, based on the equivalent
width of the H$\alpha$ line, \citet{Bouvier92} classified it as a B4
star, while \citet{Wilking05} suggested a spectral type of B3.  The star
is very bright in the near-IR and barely visible in the mid-IR, where
the emission is completely dominated by the surrounding reflection
nebula, which saturates the MIPS image close to S\,1, see
Fig.\,\ref{fig-MIPS}.  It was not detected in any of the WISE bands.
\citet{lada1984}  used  ground based near and mid-IR photometry to
determine the bolometric luminosity of S\,1. They found no IR excess for
the star after dereddening the observed flux densities with 12 mag of
visual extinction. By fitting a blackbody curve to the extinction
corrected spectral energy distribution they derived an effective
temperature, T$_{eff}$ = 16000 K, and a luminosity, L = 1100 \lsun\
(corrected for a distance of 140 pc).

Using the VizieR photometry viewer, we have compiled all available
photometric measurements (with uncertainties) of S\,1. The measurements
include values in the Johnson optical (B band) and near-infrared (JHK)
filters, POSS filters (J, F, i), Gaia G band filter, PanStarrs optical
(g, r, i) filters, SDSS optical (i$'$) filter, Elias near-infrared (J,
H, K, L$'$) filters, 2MASS near-infrared (J, H, K$_s$) filters, IRAC
infrared (3.6, 4.5, 5.8, and 8.0 $\mu$m) filters, NICMOS near-infrared
(1.10 and 1.60 $\mu$m) filters, ISOCAM (CAM1: 3.0, 3.3, and 3.7 $\mu$m;
CAM4: 5.0 - 5,7 $\mu$m). Analysis of the FORCAST images show that S\,1
is undetected in all filters. The 3-$\sigma$ upper limit at 11.1 $\mu$m
is 180 mJy. We then attempted to fit the resulting spectral energy
distribution with a blackbody function with reddening. We adopted the
\citet{Rieke85,Rieke89} extinction curve. A straightforward fit to the
data, for a fixed distance of 138 pc, but without constraints on the
reddening, radius, or temperature, does not yield results that are
consistent with either main sequence stellar parameters or previous
determinations of the extinction or spectral type for S\,1. 

\begin{figure}
\centering

\includegraphics[width=0.5\textwidth]{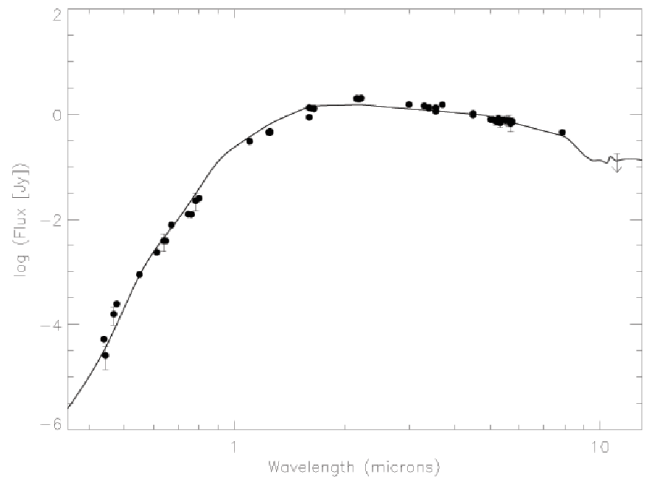}
\caption{Spectral Energy Distribution (SED) for S\,1 based on all existing photometry. The solid curve
represents the SED for a reddened blackbody corresponding to  a B4V star (T$_{eff}$ = 17000 K, R = 4.35 R$_\odot$) 
with an E$_{B-V}$ = 3.8 and R$_V$ = 3.35 (that is, A$_V = 12.7$ mag), see Section 3.
\label{fig-SED}}
\end{figure}

Because the reddening appears to be so large, the photometric data
provide relatively little power to constrain the stellar temperature.
Therefore we fixed the temperature at 17000 K, corresponding to that for
a B4V star. For consistency, we also fixed the radius to be that
appropriate for a B4V star (4.35 R$_\odot$; Schmidt-Kaler 1982). This
leaves the reddening, $E_{B-V}$, as a free parameter.  While a total
optical reddening value of A$_V$  $\sim$ 12 mag, consistent with
previous determinations, then provides a reasonable fit to the optical
data, we find that the model overestimates the photometric observations
in the near-infrared. In fact, we find that no value of the reddening
with the standard extinction law yields a model that can adequately fit
all of the photometric data. Various authors have claimed that the
reddening law towards the Ophiuchus cloud is non-standard, with values
of $R_V$ as high as 4.2 reported \citep{Chini81,Andersson07,Lee18}.
Therefore we allowed the $R_V$ value to vary along with the $E_{B-V}$
value. With the adopted stellar parameters, we find that $E_{B-V} \sim
3.8$ and $R_V \sim 3.35$ provides a good representation of all the
photometric data points from the optical out to 10 $\mu$m, and is
consistent with our measured upper limit at 11.1 $\mu$m (see Fig.
\ref{fig-SED}). These values correspond to a total extinction in the
optical of $A_V = 12.7$ mag. A slightly larger value of the total
reddening ($A_V = 13.3$) is required to fit the data if an earlier
spectral type (B3V) is adopted, but again a non-standard $R_V$ value of
3.4 is necessary. Both of these R$_V$ values are consistent with the
angular variations in R$_V$ found by \citet{Lee18} for the $\rho$\,Oph
cloud complex.\\

{\bf GSS\,30~IRS\,1:} This is a deeply embedded low-mass Class I protostar,
which illuminates a bipolar reflection nebula and drives a molecular outflow.
\citet{Je15} did a full range scan of  GSS\,30~IRS\,1 with {\it Herschel}/PACS
and found extended emission in both \OI\ and \CII. The \OI\ emission is most
likely shock excited, while the \CII\ emission is PDR emission from the
reflection nebula. \citet{Je15} estimate the  far-ultraviolet radiation field is
in the range 3 to 20 G$_0$, where G$_0$ is in units of the Habing field,
defined as $G_0$=1.6$\times$10$^{-3}$\,erg\,cm$^{-2}$\,s$^{-1}$
\citep{habing1968}.

\subsection{S\,1 PDR}

S\,1 has created a large cigar shaped ``cavity'', which may extend as
far as $\sim$ 550\arcsec\ to the east northeast, and $\sim$ 80\arcsec\
to the west southwest, where the expansion is blocked by the dense
surrounding molecular cloud, see Fig.\,\ref{fig-MIPS}. The projected
major axis of the reflection/emission nebula is therefore $\sim$ 0.6 pc.
The morphology of the S\,1 reflection nebula resembles NGC\,2023
\citep{Sandell15}, i.e., both show ellipsoidal cavities, which interact
strongly with the surrounding dense molecular cloud and expand more
freely to the opposite side of the nebula, where the density of the
surrounding cloud is much lower. As found in NGC\,2023 we see filaments
and lumpy ridges in the IRAC images, where the S\,1 reflection nebula
expands into the dense molecular cloud, suggesting that the dense
surrounding cloud is clumpy. There is a bright ridge $\sim$ 20\arcsec\
south of S\,1, which dominates the emission in the mid-IR. The \CII\
emission in our map is completely dominated by the reflection nebulosity
illuminated by S\,1, although there may be some marginal contribution
from GSS\,30. The same is true for \OI, even though one can also see
faint \OI\ emission from shock excited \OI\ in the VLA\,1623 outflow and
even stronger emission from GSS\,30.  \citep{larsson2017,Nisini15,Je15}.

\section{Results}

\begin{figure*}
\centering
\includegraphics[width=0.75\textwidth]{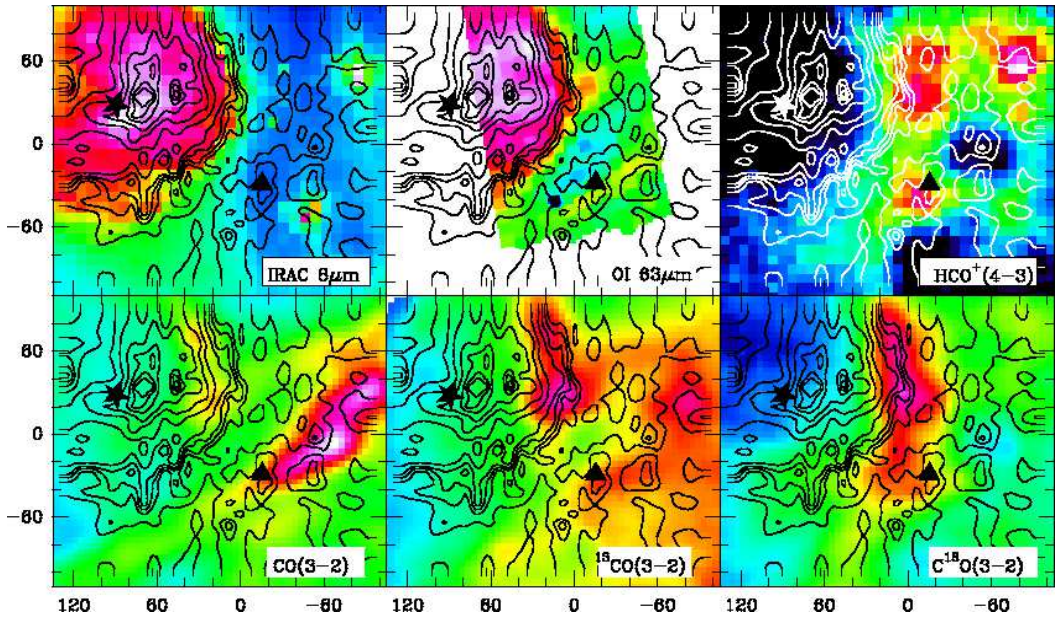}
\caption{Contours of integrated (0--10\,\kms) \CII\ intensity map of
$\rho$\,Oph\,A compared with color images of (Top,left) IRAC
8\,\,micron, (Top,middle) PACS \OI\ 63\,\micron, (Top,Right) JCMT
HCO$^+$(4--3), (Bottom, Left) CO(3--2), (Bottom, Middle) $^{13}$CO(3--2)
and (Bottom, Right) C$^{18}$O(3--2). The \CII\ contours are at 25, 40,
50, 60, 75, 100, 120, 140, 160, 170, 175 and 180\,K~\kms. The filled
star and triangle respectively, denote the positions of the star S\,1 
and the young stellar object VLA\,1623.
\label{fig_panelplot}}
\end{figure*}

Figure\,\ref{fig_panelplot} shows comparison of  measured \CII\
intensity distribution integrated over the velocity interval
0--10\,\kms\ with emission from other PDR tracers (8\,\micron\ dust
continuum and \OI\ at 63\,\micron) and from molecular material at
moderate (CO(3--2) and its isotopes) and high densities (HCO$^+$(4--3)).
The emission distribution of the PDR tracers is markedly different from
the molecular gas tracers. Among the molecular gas tracers the CO(3--2)
emission  is detected over the whole area and does not seem to trace the
PDR emission at all. The strongest CO emission comes from the blue
outflow lobe powered by the Class 0 source VLA\,1623. There is strong
HCO$^+$(4--3) emission from the dense cores in the $\rho$\,Oph\,A ridge.
Additionally, HCO$^+$ outlines the dense compressed gas layer to the
west and south, where the PDR shell is expanding into the dense
molecular cloud. The \CII\ emission is completely dominated by the
strong PDR emission from the emission/reflection nebula surrounding S\,1
with only faint emission west of the S\,1 PDR. This faint \CII\ emission
is probably mostly from the surface layers of the molecular cloud,
possibly enhanced  by FUV radiation from GSS\,30~IRS\,1. The 8\,\micron\
emission, which is dominated by PDR emission is strikingly similar to
\CII, \CII\  and 8 \micron\ emission both peak west and south of S\,1.
The \OI\ emission is also dominated by the S\,1 PDR similar to \CII.
However, \OI\ also clearly shows shock excited emission from the bipolar
outflows powered by VLA\,1623 and GSS\,30, where the \CII\ emission is
faint or completely absent. Additionally \OI\ shows a spur (protrusion)
of emission sticking out of the PDR shell approximately straight west
from S\,1. This feature appears to be real, since it is also seen in the
\OI\ map analyzed by \citet{Nisini15}. It is not evident in the
integrated \CII\ image, although there is a hint of it in the \CII\
channel maps (Fig 3) at +2 \kms. In the H$_2$ emission observed with
ISOCAM-CVF \citep{larsson2017} the hot PDR shell is completely
dominating, but especially in the S2 line we can also see additional
H$_2$ emission inside the PDR region, particularly in the southwestern
part of the PDR.

\begin{figure}
\centering
\includegraphics[width=0.50\textwidth]{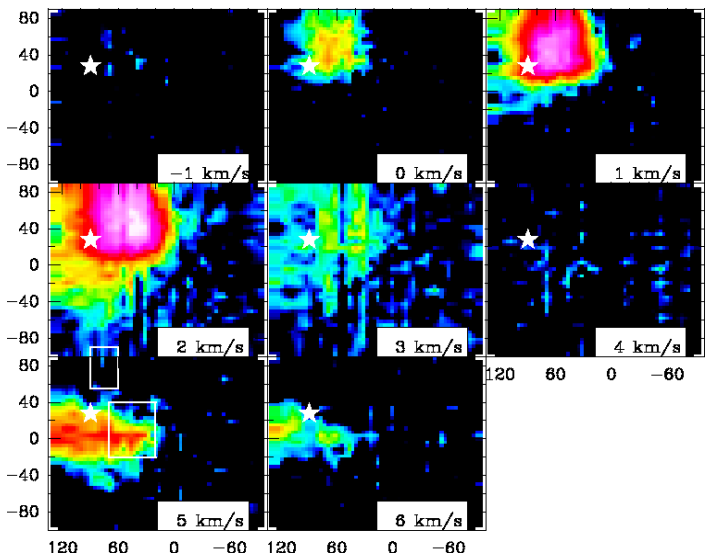}
\caption{\CII\ channel maps of 1\,\kms\ wide velocity channels spaced
by 1\,\kms. The color scale ranges from 5 to 70\,K~\kms. 
\label{fig_channelmap}}
\end{figure}

Figure\,\ref{fig_channelmap} presents channel maps of 1 \kms\ wide velocity bins
ranging from -1\,\kms\ to 6\,\kms. The maps clearly show the two main velocity
peaks, the first between 0--3\,\kms\ and the second 5--6\,\kms. Comparison of
the \CII\ channel maps with the CO (and its isotopologues) channels show that
the higher velocity component is detected only in CO(3--2) and not in the rarer
isotopologues
(Fig.\,\ref{fig_co32chan},\ref{fig_13co32chan},\ref{fig_c18o32chan}). 
HCO$^+$(4--3) is our only high density tracer, n$_{crit}$ = 2 - 3 $\times 10^6$
cm$^{-3}$ \citep{Shirley15}, and the only line, perhaps with the exception of C$^{18}$O,  which is
unaffected by self-absorption. The upper energy level, E$_u$/k =
42.8 K, however, is rather low. Therefore the emission is dominated by the dense cold
cores in the $\rho$\,Oph\,A cloud, which is seen both in the
integrated intensity map (Fig.\ref{fig_panelplot}) as well as in the channel maps
(Fig.\,\ref{fig_hcop43chan}).

\begin{figure}
\centering
\includegraphics[width=0.45\textwidth]{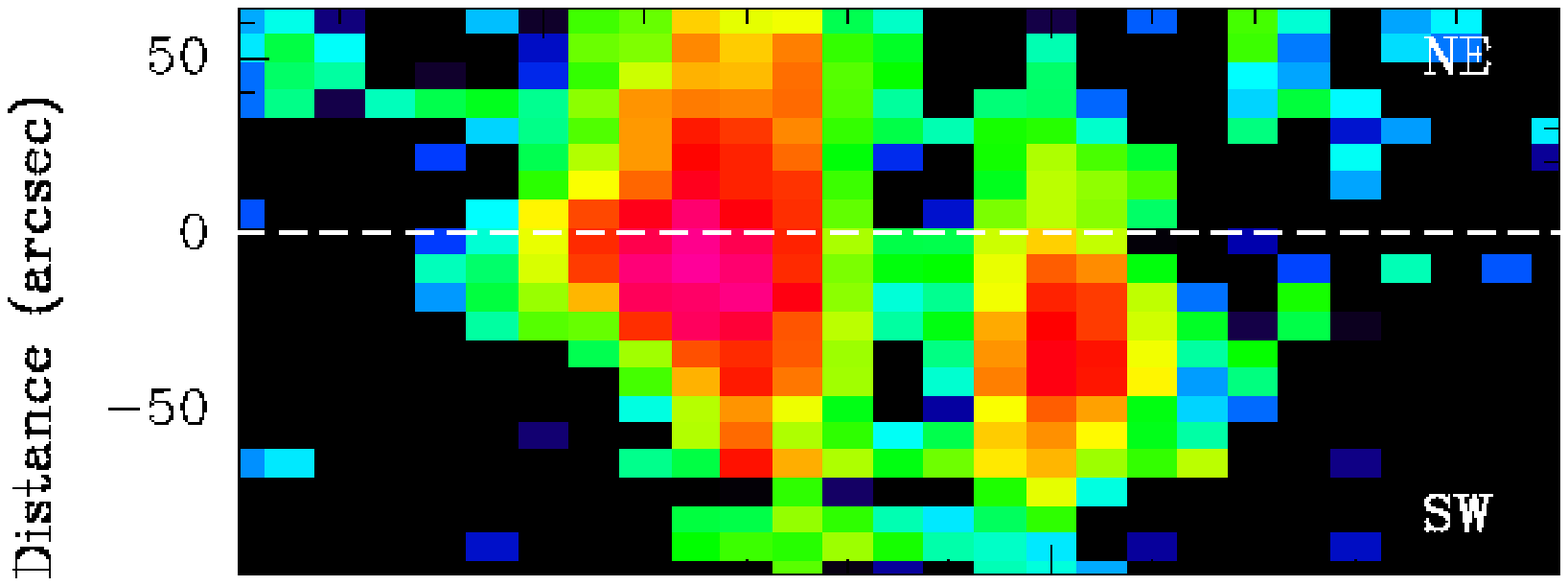}
\includegraphics[width=0.46\textwidth]{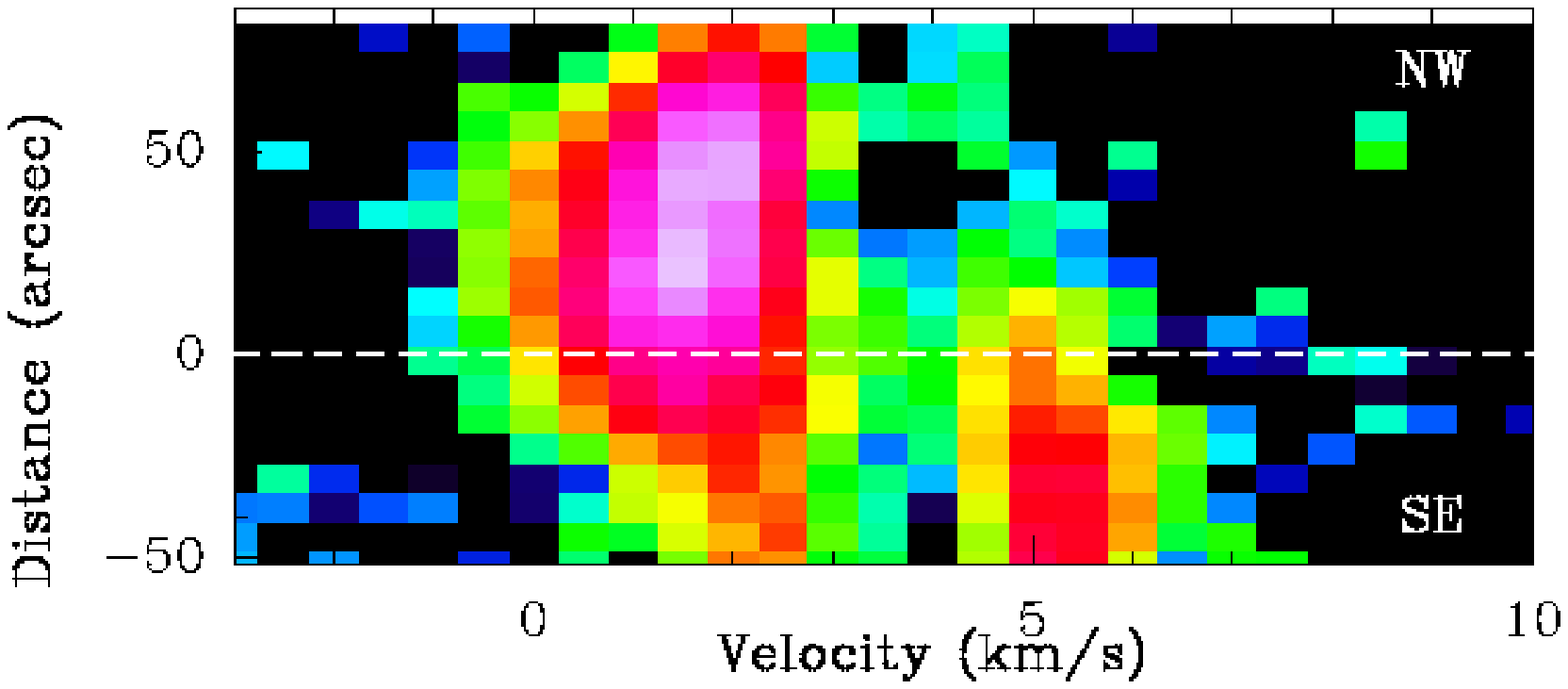}
\caption{\CII\ position-velocity diagrams at position angles of
-135\arcdeg\ (Top) and -45\arcdeg\ (Bottom). The position of S\,1
marked with the dotted line.
\label{fig_pvdiag}}
\end{figure}

The position-velocity diagrams derived for cuts at position angles of
-135\arcdeg\ and -45\arcdeg\ show that the \CII\ lines are broader
towards the center of the S\,1 nebula and narrower towards the PDR shell
(Fig.\,\ref{fig_pvdiag}).   This suggests that the \CII\ emission has
contribution from the entire C II region as well as the front and back
of the PDR shell. The \CII\ emission is dominantly blue-shifted relative
to the systemic velocity, indicating that the emission is dominated by
photo evaporation flows from the dense PDR shell. Such emission would be
blue-shifted if the densest PDR region is behind the star, i.e. the
reflection nebula is inclined towards the line of sight, so that the
large extended lobe northeast of the star is largely coming towards us.
Unfortunately our \CII\ map does not cover the eastern part of the
nebula. Therefore we cannot  confirm the proposed geometry of the
reflection nebula from the existing data sets.

\begin{figure}
\centering
\includegraphics[width=0.45\textwidth]{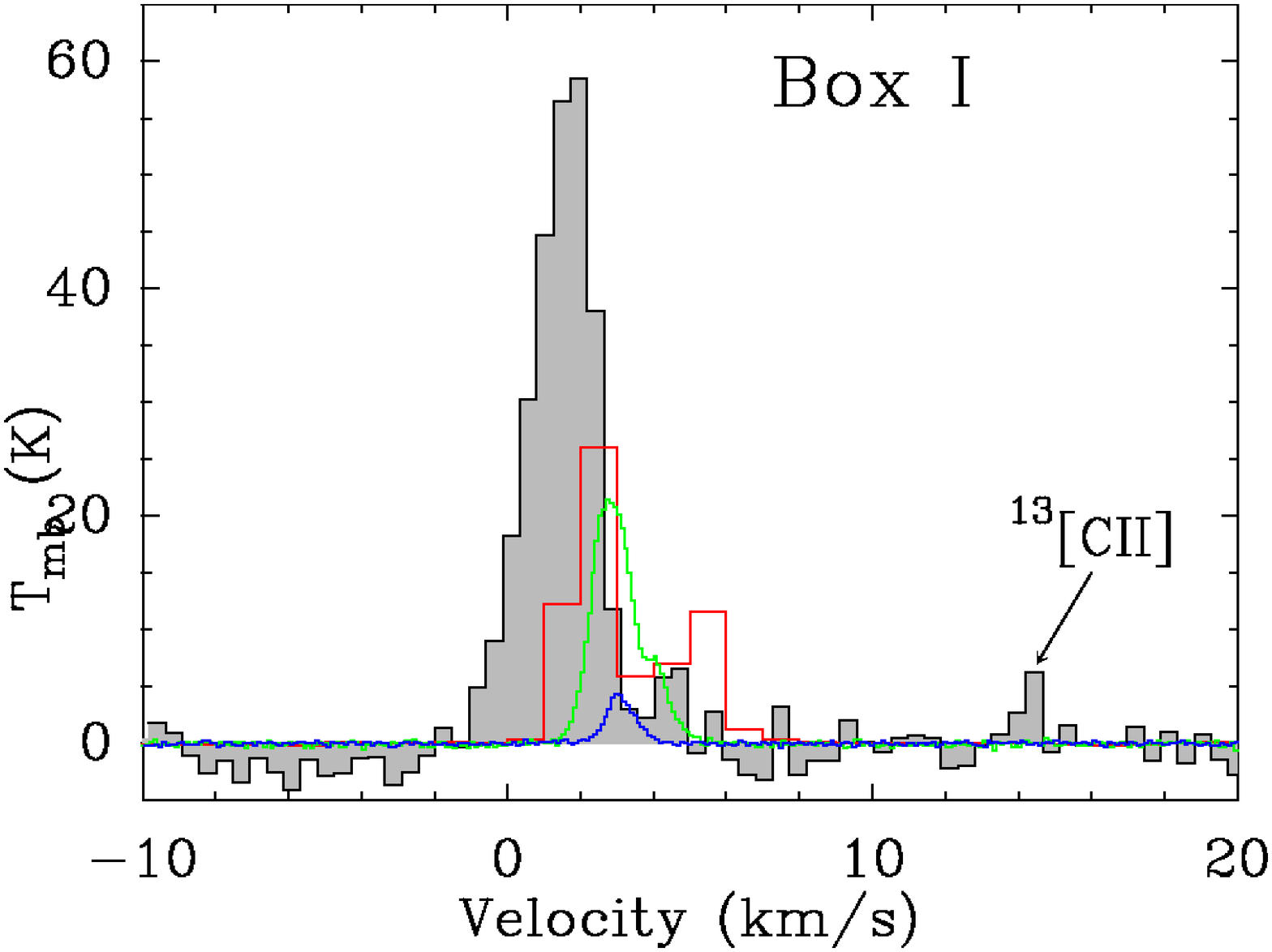}
\includegraphics[width=0.45\textwidth]{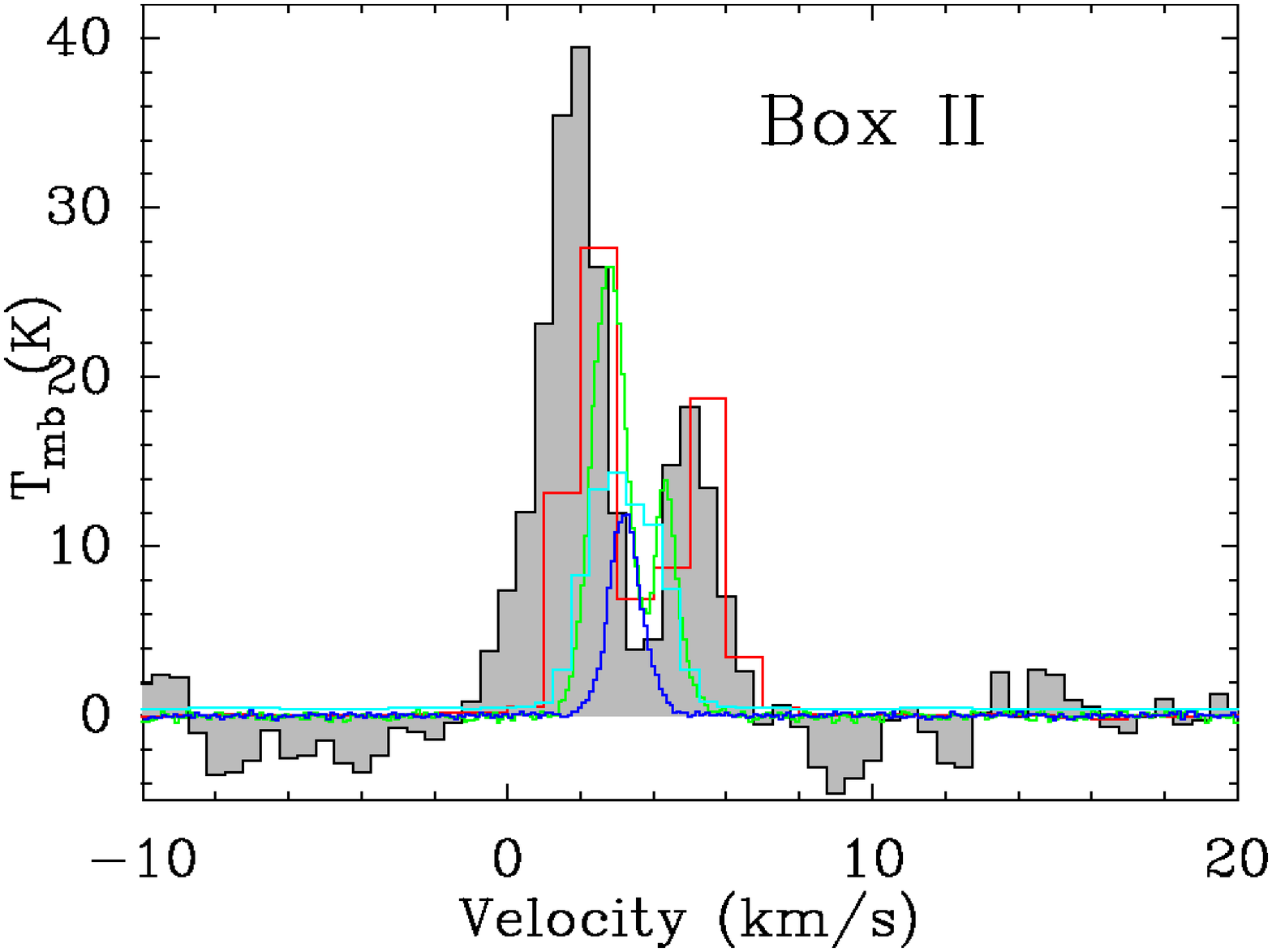}
\caption{Average spectra of \CII\ (black), CO(3--2) (red),
$^{13}$CO(3--2) (green) and C$^{18}$O(3--2) (blue) over  the regions I
and II marked in Fig.\,\ref{fig_channelmap}.
\label{fig_avgcomp}} 
\end{figure}

Figure\,\ref{fig_avgcomp} show the \CII\ spectrum averaged over the two
regions shown in Fig.\,\ref{fig_channelmap}, compared with the average
spectra of  CO and its isotopes over the same region. We find that in
Box 1, located to the north, only CO(3--2) and \CII\  show double peaked
spectra, while there is a hint of a shoulder in  $^{13}$CO(3--2).
C$^{18}$O(3--2) shows a single-peaked spectrum. The strongest hyperfine
component of \thCII, i.e. F = 2--1, red-shifted by 11.2 \kms, is also
detected with a peak temperature of $\sim 8.5$\,K, suggesting that the
\CII\ emission is optically thick in most of the region. A Gaussian fit
of the  \thCII\ component gives a V$_{\rm LSR}$ = 3.1 \kms\ (relative to
the rest frequency of the F = 2--1 line). We get exactly the same
velocity for C$^{18}$O, confirming that both CO and \CII\ are strongly
self-absorbed.  The ratio of the integrated intensities of \CII\ and
$^{13}$\CII\ F = 2--1 is $\sim 27$. The intensity of the F = 2--1
transition of \thCII\ is 0.625 of the total \thCII\ intensity.  Thus,
the ratio of intensities of \CII\ to \thCII\ is $\sim 17$, which
corresponds to an optical depth of 3.7 for \CII. Further based on the
two-component LTE model (Table\,1) for the emission (background)
component we had derived N(\cplus) = 5.6$\times$10$^{18}$\,\cmsq. This
N(\cplus) corresponds to an optical depths of 1.6 and 6, for
temperatures of 300\,K (typical PDR temperature) and 80\,K ($T_{\rm ex}$
derived from the LTE analysis) respectively.  This confirms that \CII\
is optically thick and it is not possible to discern any depletion of
$^{13}$\cplus\ due to fractionation. For Box 2, which is south of S\,1,
the self-absorption, is even more striking and seen in $^{13}$CO as
well.  Here there is only a hint of the  \thCII\ F = 2--1 hyperfine
component, because some of the spectra had rather poor baselines.
However, as in Box 1, C$^{18}$O peaks in the middle of the
self-absorption confirming that double peaked spectra are caused by
self-absorption from cold foreground gas.  We discuss the foreground
absorption in more detail in Sec.\,5.

\section{The self-absorbing \cplus\ cloud}

\begin{figure}
\centering
\includegraphics[width=0.48\textwidth]{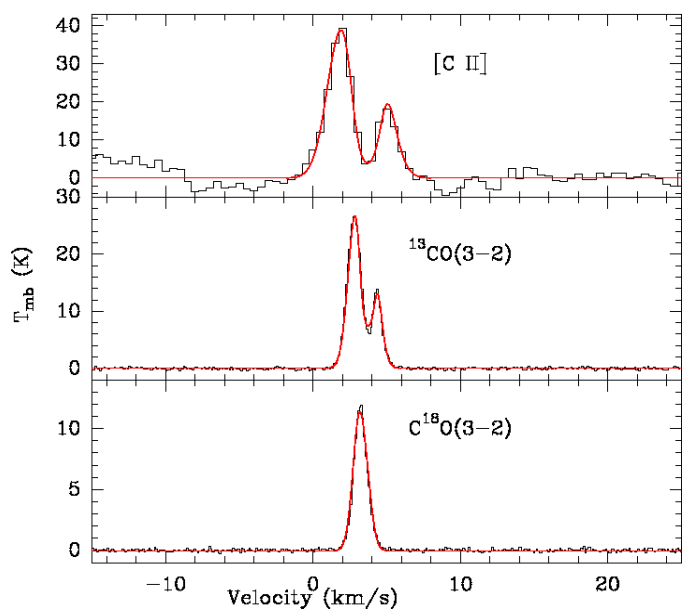}
\caption{Two component LTE model fits (in red) to average spectra
obtained using CASSIS. The fitted parameters are presented in
Table\,\ref{tab_ltemodel}.
\label{fig_ltemodel}}
\end{figure}

The available velocity resolution of the \CII\ data has enabled the
detection of strongly self-absorbed spectral profiles over an extended
region in the S\,1 PDR of $\rho$\,Oph\,A. This absoprtion dip is consistent with the
results of spectrally unresolved observations of the two fine-structure
transitions of \OI\ at 63 and 145\,\micron, which had also suggested the
presence of a foreground absorbing cloud. Here, we take advantage of the
available velocity information and characterize the background and
foreground cloud using a two-layer Local Thermodynamic Equilibrium (LTE)
model.  We used the CASSIS\footnote{http://cassis.irap.omp.eu} software
\citep{vastel2015},  to model the average spectral profiles of \CII,
$^{13}$CO(3--2) and C$^{18}$O(3--2) assuming a warm background and a
colder foreground component (Fig.\,\ref{fig_ltemodel}). CASSIS allows a
combined fit of the two components to the observed spectrum of a species
with the following adjustable parameters: column density ($N$),
excitation temperature ($T_{\rm ex}$), $\upsilon_{\rm LSR}$, FWHM of the
line ($\Delta\upsilon$) and size of the source ($\theta$). We have
assumed both components to be Gaussians. The results of the fit are
presented in Table\,\ref{tab_ltemodel}. We find that the relative
abundance of the C$^{18}$O and $^{13}$CO molecules in the two components
are significantly different. For the warmer component the ratio is $\sim
8.8$, while it is $\sim 14$ for the colder component, showing that
$^{13}$CO is significantly optically thick both in the foreground and
the background cloud.  We can convert the derived C$^{18}$O column
densities (Table 1) to total column densities using the empirical
relationship derived by \citet{Frerking82} for  $\rho$\,Oph:

\[ N(H_2) =\left[ \frac{{\rm C}^{18}{\rm O}}{1.7 \times 10^{14}} +
3.9\right]
\times 10^{21}$  cm$^{-2}\]

\noindent
which gives us 9.3 $\times$ 10$^{21}$ \cmsq\ and 5.6 $\times$ 10$^{22}$ \cmsq\
for the foreground and background cloud, respectively. If we take the derived
foreground column density and transform it to visual extinction assuming normal
gas to dust ratio, i.e.,  $N($H$_2$)/A$_{\rm V}$ = 0.94 $\times$ 10$^{21}$
molecules~cm$^{-2}$~mag$^{-1}$ \citep{bohlin1978,Frerking82}, we get an
extinction of 9.9 mag. This is lower than the estimated reddening towards S\,1, $A_{\rm V}$ = 12.7 mag, see
Sect. 3. This is to be expected, since C$^{18}$O does not
trace the hot gas in the PDR. We can get a rough estimate for the column density
of the gas in the PDR by using the derived $N$(\cplus) for the foreground \CII\
(Table 1). If we assume that most of the carbon is ionized, i.e.  [C/H] =
10$^{-4}$, we get $N($H$_2$) = 5.3\,10$^{21}$\,\cmsq, or  5.6 mag. This is
almost certainly an overestimate, since some of the carbon is likely to
be neutral or tied up in molecules like CO.

\begin{table}[h]
\tiny{
\caption{Results of fitting of two-component LTE models to average spectra
corresponding to the double-peaked profiles \label{tab_ltemodel}}
\begin{tabular}{lrrrrr}
\hline
\hline
Species &  $\upsilon_{\rm LSR}$ & $\Delta\upsilon$ & $T_{\rm
ex}$ & $N$ & $\theta$ \\
\hline
& \kms & \kms & K & \cmsq & \arcsec\\
\hline
\cplus\ & 3.1$\pm$0.1 & 2.7$\pm$0.2 & 80.8$\pm$7.0 & (5.6$\pm$2.3)\,10$^{18}$ & 50\\
        & 3.7$\pm$0.1 & 1.6$\pm$0.1 & 10.5$\pm$0.3 & (5.3$\pm$0.8)\,10$^{17}$ & 30\\
$^{13}$CO & 3.4$\pm$0.1 & 1.5$\pm$0.1 & 66.3$\pm$10.5 & (7.5$\pm$0.1)\,10$^{16}$ & 43$\pm$5\\
        & 3.7$\pm$0.1 & 0.9$\pm$0.1 & 6.8$\pm$1.3 & (1.3$\pm$0.1)\,10$^{16}$ & 31.2$\pm$2.1\\
C$^{18}$O & 3.3$\pm$0.1 & 1.1$\pm$0.1 & 70 & (8.9$\pm$0.5)\,10$^{15}$ & 50\\
        & 3.6$\pm$0.1 & 0.8$\pm$0.1 & 12 & (9.2$\pm$3.3)\,10$^{14}$ & 30.1$\pm$0.1\\
\hline
\hline
\end{tabular}
}
\end{table}

\section{Discussion}

\CII\ and \OI\ are some of the best tracers of PDRs
\citep{Hollenbach99}, since they cover a large range of physical
condition present in many PDRs.  The \CII\ 158 \micron\ line, assumed to
be largely optically thin,  can be used to determine the total mass of
the emitting gas. However, with the advent of the high spectral
resolution spectrometers like Herschel/HIFI and SOFIA/GREAT there has
been a paradigm shift in our understanding of the abundance and ubiquity
of \cplus. Not only was the \CII\ line found to be optically thick,
a fact often clearly substantiated by the detection of  hyperfine lines
of $^{13}$\cplus\ \citep{Ossenkopf2013}, but a significant fraction of
\cplus\ emission was found to arise from the ISM where no corresponding
CO emission was detected in both our Galaxy \citep[GOTC+][]{langer2010}
and external galaxies \citep{mookerjea2016}.

Our \CII\ observations of $\rho$\,Oph-A with SOFIA/GREAT has also
provided useful insight into the distribution of the PDR material,
particularly in the region around the B-type star S\,1.  Unlike the
emission from CO (and its isotopes) and to some extent \OI\  the \CII\
emission is completely dominated by the emission/reflection nebula
illuminated by  S\,1.  This suggests that the observed \CII\ emission is
primarily due to the FUV radiation from S\,1 with a negligible
contribution from HD\,17889.  The PDR is a slightly asymmetric sphere
which is pressure-bound to the south and west due to the enhanced
density of the dense $\rho$\,Oph-A molecular cloud and has an average
radius of 80\arcsec. It almost certainly extends much further to the
northeast (see Fig. 1), where the \CII\ emission has not yet been
observed. 

Because we only have velocity resolved \cplus\ spectra of the PDR
emitted region, the two-component (emission and absorption) model fitted
to the \CII\ emission is not well constrained. There is a possible
degeneracy between the column density of \cplus\ and the size of the
emission region. Our current model shows consistent excitation
temperatures and sizes for \CII\ and the CO lines, although owing to the
complication of the presence of the foreground absorption, the intrinsic
\CII\ emission is not well constrained. \citet{graf2012} detected
similar optically thick \CII\ lines affected by strong self-absorption
in NGC\,2024 and estimated a very high value of $N$(\cplus)= 1.6$\times
10^{19}$\,\cmsq.  For the foreground cloud in NGC\,2024 these authors
estimated $N$(\cplus)$\geq 10^{18}$\,\cmsq\ for a gas temperature above
40\,K, but this is a very extreme region.  The $N$(\cplus) we obtain for
the foreground and background clouds in the S\,1  PDR are about 50\% of
the column densities estimated for NGC\,2024 and the temperatures we
estimate are also much lower. Using the stellar parameters for S\,1
\citet{larsson2017} estimated the FUV flux in the PDR to be 5000\,G$_0$.
In another method we first estimated the total far-infrared (FIR)
intensity from the PACS 70 and 160\,\micron\ flux densities
(Fig\,\ref{fig_fuv}) following the method described by
\citet{roccatagliata2013}. Next we assumed that the FUV energy absorbed
by the grains is re-radiated in the far-infrared \citep{mookerjea2011}
and estimated the FUV flux, $I_{\rm FUV}$ (6\,eV $<$h$\nu$ $<$ 13.6 eV)
impinging onto the cloud surfaces from the emergent FIR intensities
($I_{\rm FIR}$) using $I_{\rm FUV}$ (expressed in units of G$_0$)=
4$\pi$ $\frac{I_{\rm FIR}}{G_0}$.  From the second approach, we find the
FUV intensity at the position $\Delta \alpha$ = +65\arcsec,
$\Delta\delta$ = +58\arcsec\ to be 3100\,G$_0$. The [OI(145)]/[OI(63)]
ratio at this position is 0.15, which is much larger than the typical
ratio of 0.05--0.1 found in PDRs when both lines are optically thin.
\citet{larsson2017} explained the high values of [OI(145)]/[OI(63)]
ratio to be due to \OI(63) being partially absorbed by a cold foreground
cloud, while \OI(145) remained unaffected. Our observations support this
conjecture. The OI(145) emission, however, should mostly be optically
thin.  We can therefore use the \OI(145)]/\CII\ ratio and compare it to
predictions of the plane-parallel PDR model by \citet{kaufman2006}. At
this position the \OI(145)]/\CII\ ratio is 0.27 (in energy units), which
for the FUV intensities of 3000 and 5000\,G$_0$ correspond to the
densities of 4$\times 10^3$\,\cmcub\ and 3$\times 10^3$\,\cmcub\
respectively.  This matches very well with the critical density of the
158\,\micron\ transition of \CII.  The critical density of the \OI\
145\,\micron\ transition however is $> 10^6$\,\cmcub, suggesting that
\OI\ mostly comes from high density gas compressed between the \CII\
emitting gas and the surrounding dense molecular cloud. For the above
mentioned results the plane-parallel PDR model predicts an integrated
intensity of 100\,K~\kms\ for the \CII\ line. The background component
for the two-component LTE model fitted to the observed spectrum
corresponds to an integrated intensity of 170\,K~\kms.  This suggests
the existence of multiple \CII\ emitting PDR surfaces in contrast to a
single PDR slab, and is consistent with the scenario in which high
density clumps with PDR surfaces emitting both \CII\ and \OI(145) and
low density PDR gas emitting only \CII\ co-exist. 

In this analysis, we have assumed that the entire \CII\ emission
arises from neutral PDR.  However since carbon has an ionization
potential of 11.26\,eV, \cplus\ is also likely to exist in the \HII\
region, and
this will contribute to some extent to the observed \CII\ emission.
 As detailed in Sec.\,3, based
on radio observations the emission from the \HII\ region is rather
faint, 4 mJy at 5\,GHz. The diameter of the \HII\ region  $<$ 20\arcsec,
which is more than three times smaller than the strongest emission
feature observed in \CII\ and the other PDR tracers.  These indicate
that the contribution of the \HII\ region to the \CII\ emission is
likely to be insignificant, although an accurate estimate of $n_{\rm e}$
not being available direct comparison with the models by
\citet{abel2006} is not possible. An alternative method, is to compare
with the fine-structure transitions of N$^+$ at 122 and 205\,\micron,
since these lines arise only in the \HII\ region and have excitation
condition and critical density similar to that of \CII. For this purpose
we have used the GREAT/SOFIA observations of the 205\,\micron\ \NII\
spectra. We do not have any detection of the 205\,\micron\ line in the
entire observed region good to an r.m.s. of integrated intensity of
0.7\,K\,\kms.  \citet{abel2006} showed that the \CII\ intensities
originating from \HII\ regions and the intensities of the \NII\
205\,\micron\ line are tightly correlated following the relation ${\rm
\log I^{CII}_{H^+} = 0.937\log I^{NII}_{H^+}+0.689}$ (erg cm$^{-2}$
s$^{-1}$).  Considering the \NII\ (205\,\micron) beamsize of 21\arcsec,
for the upper limit of \NII\ intensity being 0.7\,K\,\kms, we obtain ${\rm
I^{CII}_{H^+}\sim 22}$\,K\,\kms. The peak integrated intensity of the
observed \CII\ map is 183\,K\,\kms\ and goes down to about 130\,\kms\ in a
radius of 20\arcsec\ (upper limit on the size of the \HII\ region) from
the peak position. Thus we derive an upper limit of 12-17\% for the
contribution of the \HII\ region to the \CII\ emission. Given that the
contribution is low and is also an upper limit, we conclude that our
assumption that the entire \CII\ arises from the neutral PDR does not
affect the outcome of PDR models.

\section{Summary \& Conclusion}

We have used the velocity information of the \CII\ data observed with
SOFIA/GREAT to obtain a cohesive picture of the geometry and
illumination of the PDR around S\,1 in $\rho$\,Ophiuchus. The \CII\
emission appears to arise from both photo-evaporated flow from the dense
PDR shell as well as from ionized gas inside the PDR shell. The \CII\
emitting PDR is of moderate density ($\sim 10^3$\,\cmcub), while the 63
and 145\,\micron\ emission of \OI\ requires much higher densities, i.e.,
it is mostly coming from the compressed neutral layer between the
ionized C$^+$ gas and the dense surrounding cloud as well as from dense
clumps embedded in the diffuse ionized gas.  The \CII\ map was observed
during the Early Science phase of SOFIA and does not cover the nebula to
the north-east, which prohibits us from fully constraining the geometry
of the C {\sc ii} region excited by S\,1. With the Low Frequency Array
(LFA) \citep{Risacher16}, which is a 2 $\times$ 7 pixel LFA array, where
each pixel is about four times as sensitive as the L2 receiver used in
Early Science, one can map the whole S\,1 reflection nebula with
significantly higher S/N in a fraction of the time it took to obtain the
\CII\ map discussed here. Since the LFA array is operated in parallel
with the High Frequency Array (HFA) in upGREAT, one will additionally
get velocity resolved \OI\ 63 \micron\ the same time.  The only thing
missing to fully characterize the PDR emitting regions is high-J CO
lines, which will be quite strong in the PDR shell. We know from our
work on NGC\,2023 \citep{Sandell15} that the hot molecular gas in the
PDR starts to dominate from J = 6--5 upwards.  CO(6--5), (7--6), and
(8--7) can be observed with ground based telescopes, while upGREAT on
SOFIA can observe higher J transitions starting from CO(11--10). To
accurately determine the density and temperature of the hot molecular
PDR emission we therefore need follow up observations of high-J CO
transitions both with ground based telescopes and SOFIA.

\begin{acknowledgements}
Based on observations made with the NASA/DLR Stratospheric Observatory for
Infrared Astronomy (SOFIA). SOFIA is jointly operated by the Universities Space
Research Association, Inc. (USRA), under NASA contract NAS2-97001, and the
Deutsches SOFIA Institut (DSI) under DLR contract 50 OK 0901 to the University
of Stuttgart. The development of GREAT was financed by the participating
institutes, by the Federal Ministry of Economics and Technology via the German
Space Agency (DLR) under Grants 50 OK 1102, 50 OK 1103 and 50 OK 1104 and within
the Collaborative Research Centre 956, sub-projects D2 and D3, funded by the
Deutsche Forschungsgemeinschaft (DFG). This research has made use of the VizieR catalogue access tool, CDS,
 Strasbourg, France. The original description of the VizieR service was
 published in A\&AS 143, 23.

\end{acknowledgements}

\begin{appendix}

\section{PACS continuum images and derived FUV intensity map}
\begin{figure}[h]
\centering
\includegraphics[width=0.48\textwidth]{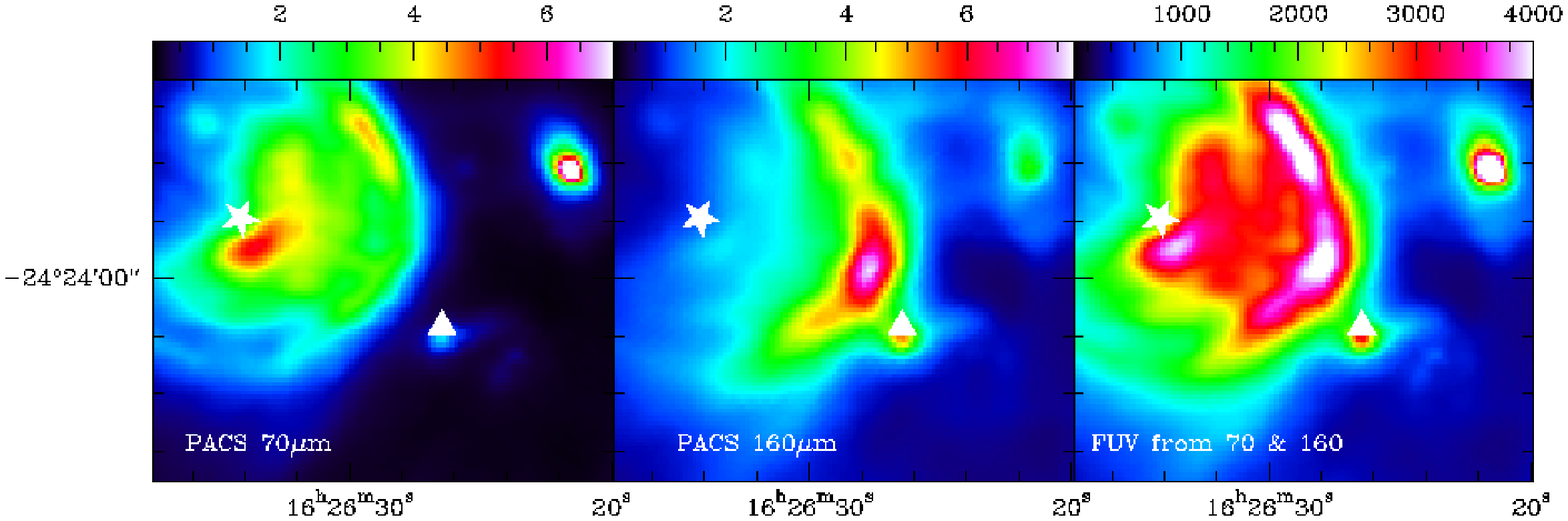}
\caption{PACS continuum maps at 70 (left) and 160 (middle) \micron,
both at the resolution of the 160\,\micron\ map. (Right) FUV intensity
derived from the total FIR intensity calculated from the PACS
70 and 160\,\micron\ continuum intensities using the method described in
text.
\label{fig_fuv}}
\end{figure}

\section{Channel Maps for Molecular Tracers}
\begin{figure}[h]
\centering
\includegraphics[width=0.45\textwidth]{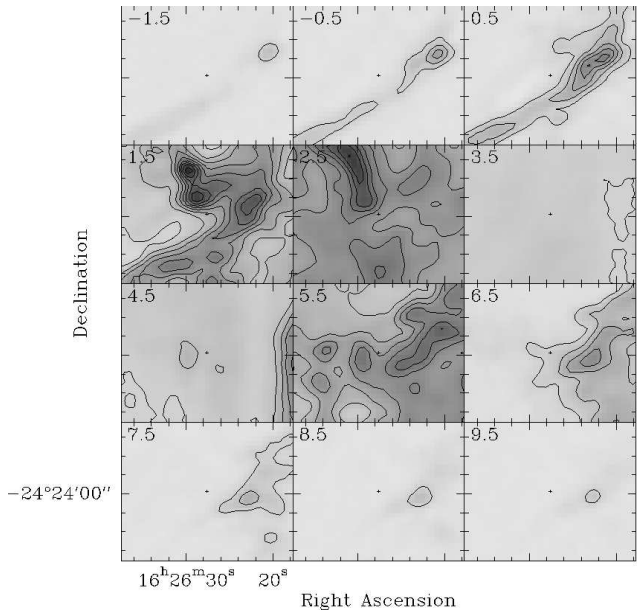}
\caption{CO(3--2) channel maps of 1\,\kms\ wide velocity channels spaced
by 1\,\kms. \label{fig_co32chan}}
\end{figure}

\begin{figure}[h]
\centering
\includegraphics[width=0.45\textwidth]{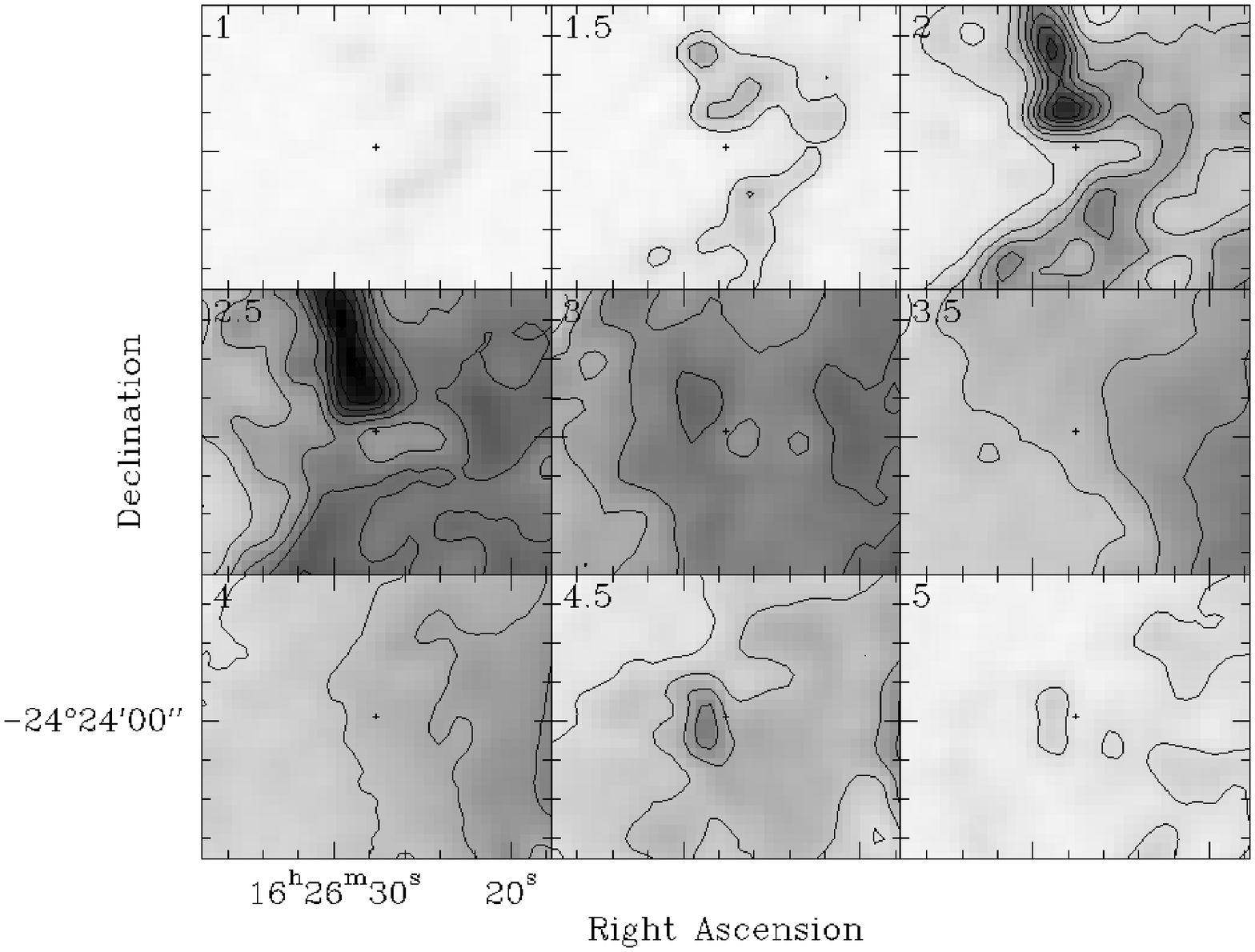}
\caption{$^{13}$CO(3--2) channel maps of 1\,\kms\ wide velocity channels spaced
by 1\,\kms. \label{fig_13co32chan}}
\end{figure}

\begin{figure}[h]
\centering
\includegraphics[width=0.45\textwidth]{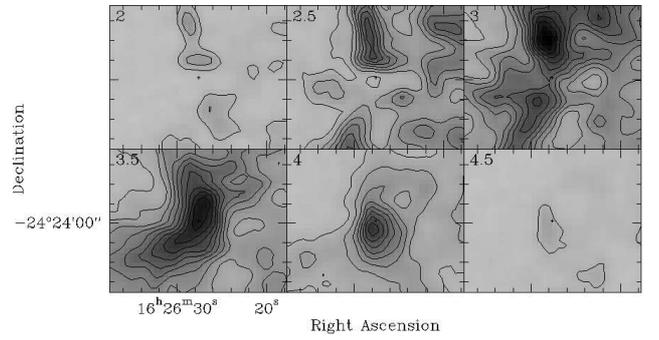}
\caption{C$^{18}$O(3--2) channel maps of 1\,\kms\ wide velocity channels spaced
by 1\,\kms. \label{fig_c18o32chan}}
\end{figure}
\begin{figure}[h]
\centering
\includegraphics[width=0.45\textwidth]{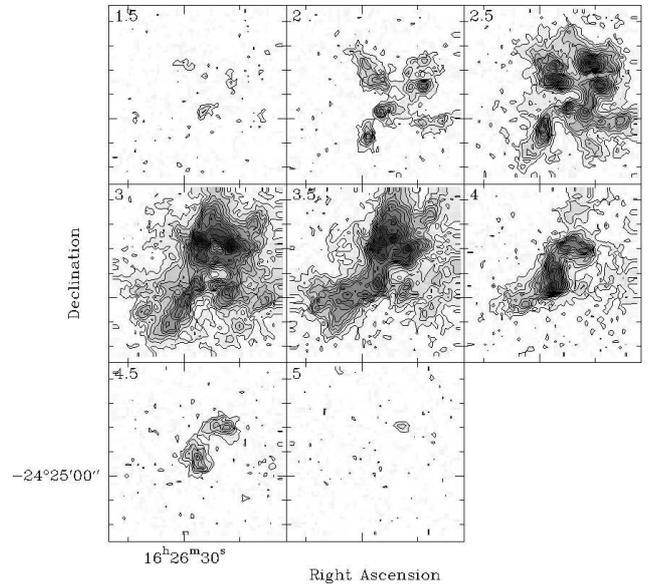}
\caption{HCO$^{+}$(4--3) channel maps of 0.5\,\kms\ wide velocity channels 
spaced by 0.5\,\kms. \label{fig_hcop43chan}}
\end{figure}

\end{appendix}

\end{document}